\documentclass{article}
\usepackage{amsthm}
\usepackage{amssymb}
\usepackage{amsmath}
\usepackage{amsfonts}
\usepackage{color}
\usepackage{indentfirst} 
\usepackage{graphicx}
\usepackage[square, comma, sort&compress, numbers]{natbib}

\newtheorem{remark}{Remark}[section]

\newtheorem{lemma}{Lemma}[section]
\newtheorem{theorem}{Theorem}[section]

\newtheorem{corollary}{Corollary}[section]
\newtheorem{definition}{Definition}[section]

\def\b1{\mbox{\boldmath $1$}}

\newenvironment{demo*}{\vspace{3mm}\noindent{\bf Proof.}}{\hfill $\Box$ \vspace{3mm}}

\begin{document}
\title{\bf \Large {A unifying approach to constrained and unconstrained \\optimal reinsurance}}
{\color{red}{\author{
\normalsize{Yuxia Huang\;\; Chuancun Yin}\\
{\normalsize\it  School of Statistics,  Qufu Normal University}\\
\noindent{\normalsize\it Shandong 273165, China}\\
e-mail:  ccyin@mail.qfnu.edu.cn}}}
\maketitle
\vskip0.01cm
\centerline{\large {\bf Abstract}}
\indent In this paper, we study two classes of optimal reinsurance models from perspectives of both insurers and reinsurers by minimizing their convex combination where the risk is measured by a distortion risk measure and the premium is given by a distortion premium principle. Firstly, we show that how optimal reinsurance models for the unconstrained optimization problem and constrained optimization problems can be formulated in a unified way. Secondly, we propose a geometric approach to solve optimal reinsurance problems directly. This paper considers a class of increasing convex ceded loss functions and derives the explicit solutions of the optimal reinsurance which can be in forms of quota-share, stop-loss, change-loss, the combination of quota-share and change-loss or the combination of change-loss and change-loss with different retentions. Finally, we consider two specific cases: Value at Risk (VaR) and Tail Value at Risk (TVaR).

\medskip

\numberwithin{equation}{section}
\section{Introduction}\label{intro}
\indent{Reinsurance is an effective risk management tool for an insurance company. By balancing paid loss and reinsurance premium, the insurance company can control its risk by sharing a portion of loss. Let $X$ be the initial loss by an insurer. Assuming $X$ is a non-negative random variable with cumulative distribution function ${F_{X}(x) = P(X\leq x)},$ survival function ${S_{X}(x) = P(X>x)}$ and $0 < E[X] < \infty.$ In order to avoid serious claims, the insurance company purchases reinsurance from another company and pay a certain amount of expense to the reinsurance company as compensation. Suppose the insurance company separates the loss function as ${f(X)}$, $0 \leq f(X) \leq X$, then the insurer retains the loss ${X-f(X)}$, denoted as ${I_{f}(X)}$.
Let ${\Pi_{f}(X)}$ denotes the reinsurance premium which corresponds to a ceded loss function ${f(X)}$, ${T_{I_{f}}(X)}$ and ${T_{R_{f}}(X)}$ represent the total loss of the insurance company and the reinsurance company, respectively.} Then we obtain the following relationships:
\begin{equation}
{T_{I_{f}}(X)}= {I_{f}(X)}+ {\Pi_{f}(X)}
\end{equation}
and
\begin{equation}
{T_{R_{f}}(X)}=f(X)-{\Pi_{f}(X)}.
\end{equation}
Let $ T(X)$ represents the convex combination of the total loss of the insurer and the reinsurer, as follows:
\begin{equation}
{T(X)=\beta T_{I_{f}}(X)+(1-\beta)T_{R_{f}}(X)},\quad \beta \in [0,1].
\end{equation}
\indent{The development of optimal reinsurance has gone through a long period of time. Borch (1960) demonstrated that the stop-loss reinsurance is optimal when the risk of the insurer is measured by variance under the expected value principle. Arrow (1963) showed that the stop-loss reinsurance is optimal when the insurer is an expected utility maximizer under the expected value principle. These basic conclusions have been extended to a number of interesting and important directions. For example, Young (1999), Kaluszka (2001), Kaluszka and Okolewski (2008). Cai and Tan (2007) proposed two optimization criterion that minimize total loss of the insurer by the Value at Risk (VaR) and the Conditional Tail Expectation (CTE). Cai et al. (2008) showed that quota-share and stop-loss reinsurance are optimal when they studied a class of increasing convex ceded loss functions by VaR and CTE under the expected value principle. Cheung (2010), Tan et al. (2011), Chi and Tan (2011), Chi and Tan (2013), Li et al. (2015) extended the fundamental results. Cheung et al. (2014) extended the conclusion of Tan et al. (2011) to the general convex risk measure that satisfied regular invariance. There are many studies about the distortion risk measures and orders, for example, Yin and Zhu (2016), Yin (2018). Chi and Tan (2013), Chi and Weng (2013) studied a class of premium principles which preserve the convex order. Zheng and Cui (2014) discussed the general model of the distortion risk measure and assumed that the distortion function is piecewise convex or concave. Cui et al. (2013) studied the general model with distortion risk measures and distortion premium principles. Cheung and Lo (2015) expended the model of Cui et al. (2013) under the cost-benefit framework. Assa (2015) studied the optimal reinsurance model of Cui et al. (2013) without the premium constraint by a marginal indemnification function (MIF) formula. Zhuang et al. (2016) studied the optimal reinsurance with premium constraint by combining the MIF formula and the Lagrangian dual method. Jiang et al. (2017) studied the Pareto-optimal reinsurance with risk constraints under the distortion risk measure. Motivated by Cai et al. (2008), Zhuang et al. (2016) and Jiang et al. (2017), we want to seek a unified way to solve this class of constrained optimization problems.}\\
\indent{The rest of the paper is organized as follows. In Section 2, we give two classes of optimal models: unconstrained and constrained optimal models. Moreover, we propose a unified framework for two classes of problems. In Section 3, we give a geometric approach to solve the object function and derive the optimal reinsurance. In Section 4, we give numerical examples about VaR and TVaR. Section 5 concludes this paper.}\\
 \section{The model}
\setcounter{equation}{0}
\indent In this section, we set up the optimal reinsurance model for the insurer and the reinsurer. Moreover, we propose a unified way to solve the unconstrained optimization problem and constrained optimization problems. We start this section from giving a brief description of the distortion risk measure and premium.
 \subsection{Distortion risk measure and premium}
\setcounter{equation}{0}
\indent{Throughout the paper, we define VaR and TVaR as ${\rm VaR}_{\alpha}(X)=\inf{\{x:P(X>x)\leq\alpha\}}$ and ${\rm TVaR}_{\alpha}(X)\\=\mathbb{E}[X\mid X\geq {\rm VaR}_{\alpha}(X)]$.}
 \begin{definition}
 A distortion risk measure of a non-negative random variable $X$ is defined as
\begin{equation}
{\varrho_{g}(X)={\int_{0}^\infty}g(S_{X}(x))dx},
\end{equation}
where function ${g:[0,1]\rightarrow[0,1]}$ is non-decreasing, g(0)=0 and g(1)=1.
\end{definition}
\indent From the Fubini theorem, we have
\begin{equation}
\varrho_{g}(X)={\int_{0}^1}VaR_{\alpha}(X)dg(\alpha).
\end{equation}
\begin{definition} A distortion premium principle of a non-negative random variable $X$ is defined as
 $$ {\Pi(X)}=(1+{\rho}){\varrho_{g}(X)},$$ where $\rho>0$ is the safety loading.
\end{definition}
\indent We achieve the expression of reinsurance premium $ {\Pi_{f}(X)}$ which corresponds to a ceded loss function ${f(X)}$ as follows:
\begin{equation}
{\Pi_{f}(X)}=(1+{\rho}){\varrho_{g}(f(X))}.
\end{equation}
\indent{If $g(x)=x$, then the distortion premium principle recovers the expected value principle. If the distortion function is concave with $\rho=0,$ then the distortion premium principle recovers Wang's premium principle.}
\begin{remark}
\indent In this paper, assuming the confidence level of a distortion risk measure is $1-\alpha$ $(0<\alpha<1)$ and the confidence level of a distortion premium principle is $1-\gamma$ $(0<\gamma<1)$.
For the convenience of discussion, we give the following definition
\begin{equation}
K(t) \triangleq \frac{g_{\alpha}(t)}{g_{\gamma}(t)},\quad t\in(0,1).
\end{equation}
\indent{Note that $K(t)$ may be convex, concave, piecewise convex or concave, where we only discuss the case that $K(t)$ is a concave function. Other cases can take the same method to discuss.}
\end{remark}
\indent{In the following subsection, we will start from two optimization problems to study two classes of optimization problems.}
\subsection{Model setup}
\setcounter{equation}{4}
\setlength{\parindent}{1em}\indent{Let $\cal{H}$ denote a class of ceded loss functions, which consist of all $h(x)$ defined on [0,$\infty$) with the form $$h(x)=\sum_{j=1}^n C_{n,j} ( x-d_{n,j}) _{+},{\ x\geq0},  \ n=1,2, \cdot\cdot\cdot,$$
where ${C_{n,j}} \geq 0$ and ${d_{n,j}} \geq 0$ are constants such that $0 {\leq} {\sum_{j=1}^n C_{n,j}} {\leq} 1 ,$ ${d_{n,1} \leq d_{n,2} \leq \cdot\cdot\cdot \leq d_{n,n}}$.}\\
\indent{Let $\cal{F}$ = \{$f(x)$: $f(x)$ is increasing convex function with $0 \leq f(x) \leq x $ for $x\in [0,\infty)$\}. Note that $\cal{H} \subset \cal{F}$.}\\
\indent{Assuming that $f^{*}$ is an optimal reinsurance strategy, from Lemma 3.2 of Cai et al. (2008) and the Fubini Theorem, we have
\begin{equation}
{\beta}\varrho_{g_{\alpha}} {(T_{I_{f^{*}}}(X))}+(1-{\beta}) \varrho_{g_{\alpha}} {(T_{R_{f^{*}}}(X))}=\underset{h\in \cal H} {\min}\{{\beta}\varrho_{g_{\alpha}} {(T_{I_{h}}(X))}+(1-{\beta}) \varrho_{g_{\alpha}} {(T_{R_{h}}(X))}\}.
\end{equation}}
\indent{Now we consider an idealized case. The insurer and the reinsurer have enough capital, so they do not worry the loss they will bear and the insurer can pay reinsurance premium without the budget constraint when they design the reinsurance contract. In this case, we give the unconstraint optimization model as follows.}\\
{\bfseries Mode 1} ( Unconstrained optimization model )
\begin{equation}
\min\{\varrho_{g_{\alpha}}(T(X))\}.
\end{equation}
\indent In realistic insurance application, risk regulators of the insurer and the reinsurer will require that their loss be limited in a range, and the insurer will have a budget constraint for the reinsurance premium. In this case, we give the constraint optimization model as follows.\\
{\bfseries Mode 2} ( Constrained optimization model )
\begin{equation}
\left\{ \begin{aligned}   \overset{} \min\{\varrho_{g_{\alpha}}(T(X))\} ,\\ s.t. \quad  \varrho_{g_{\alpha}} ({T_{I_{h}}(X)})\leq L_{1}, \\ \varrho_{g_{\alpha}} ({T_{R_{h}}(X)})\leq L_{2},\\ \Pi_{h} (X)\leq L_{3}, \end{aligned}   \right.
\end{equation}
where $L_{1}$, $L_{2}$ and $L_{3}$ are some monetary levels.\\
\indent It is important to find a unified approach to address the unconstrained optimization problem and constrained optimization problems. In the next we will show how the optimal reinsurance design for the unconstrained optimization problem and constrained optimization problems can be formulated in the same way.\\
\indent Denoting the unconstraint optimization model as follows:
\begin{equation}
\min L^{1}_{h}(X)\triangleq \min\{\varrho_{g_{\alpha}}(T(X))\}.
\end{equation}
By the Lagrangian dual method (From Jiang et al. (2017) with the formula (16)), (2.7) can be expressed as
\begin{equation}
\min L^{2}_{h}(X)\triangleq \min\{\varrho_{g_{\alpha}}(T(X))+\lambda_{1}( \varrho_{g_{\alpha}} ({T_{I_{h}}(X)})-L_{1})+\lambda_{2}(\varrho_{g_{\alpha}} ({T_{R_{h}}(X)})- L_{2})+\lambda_{3}(\Pi_{h} (X)- L_{3})\},
\end{equation}
where $\lambda_{i} > 0$ for $i = 1, 2, 3.$ We derive that (2.8) and (2.9) can be represented in the unified form
\begin{equation}
\min L^{3}_{h}(X)\triangleq \min\{\varrho_{g_{\alpha}}(T(X))+\lambda_{1}( \varrho_{g_{\alpha}} ({T_{I_{h}}(X)})-L_{1})+\lambda_{2}(\varrho_{g_{\alpha}} ({T_{R_{h}}(X)})- L_{2})+\lambda_{3}(\Pi_{h} (X)- L_{3})\},
\end{equation}
where $\lambda_{i} \geq 0,$ $i=1,2,3.$
\\ \indent{From (2.10), we achieve the followings.}
\\ \indent{Case 1: $\lambda_{1}>0, \lambda_{2}=\lambda_{3}=0$, which mean that the insurer limits its loss in a range.}
\\ \indent{Case 2: $\lambda_{2}>0, \lambda_{1}=\lambda_{3}=0$, which mean that the reinsurer limits its loss in a range.}
\\ \indent{Case 3: $\lambda_{3}>0, \lambda_{1}=\lambda_{2}=0$, which mean that the insurer has a reinsurance premium budget constraint. For example, Zheng et al. (2014) and Zhuang et al. (2016).}
\\ \indent{Case 4: $\lambda_{1}>0, \lambda_{2}>0, \lambda_{3}=0$, which mean that two insurance companies all control their loss in a range. For example, Jiang et al. (2017).}
\\ \indent{Case 5: $\lambda_{1}>0, \lambda_{3}>0, \lambda_{2}=0$, which mean that the insurer has a loss constraint and a reinsurance premium budget constraint.}
\\ \indent{Case 6: $\lambda_{2}>0, \lambda_{3}>0, \lambda_{1}=0$, which mean that the insurer has a reinsurance premium budget constraint and the reinsurer limits its loss in a range.}
\\ \indent{We know that solving these optimal problems is transformed into solving (2.10).
In the next section, we will solve (2.10) by a geometric approach. Before that we conclude this section by introducing the following notations.} For $\beta\in[0,1]$, $\lambda_{i} \geq 0,$ $i = 1,2,3,$ we denote
\begin{equation}
{\beta}+\lambda_{1} = m_{1}, \hspace{0.5cm} 2\beta-1+\lambda_{1}-\lambda_{2} = m_{2}, \hspace{0.5cm} (1+\rho)(2\beta-1+\lambda_{1}-\lambda_{2}+\lambda_{3}) = m_{3},
\end{equation}
\begin{equation}
\lambda_{1}L_{1}+\lambda_{2}L_{2}+\lambda_{3}L_{3}=D,\hspace{0.5cm} \left.M=m_{3}\small/ m_{2} \right.,
\end{equation}
\begin{equation}
\overline{X}=\sup\{X: X\in[0,\infty)\},
\end{equation}
\begin{equation}
\overline{K}= \sup\{K(t): t\in (0,1)\}, \hspace{0.5cm} \underline{K} = \inf\{K(t): t\in (0,1)\},
\end{equation}
\begin{equation}
H(x)=m_{2}g_{\alpha}(S_{X}(x)) - m_{3}g_{\gamma}(S_{X}(x)).
\end{equation}
\vskip 0.2cm
 \section{Optimal reinsurance contract}
\setcounter{equation}{0}
\indent In this section, we will derive the solution of these optimal problems. Now, we give the specific expression of (2.10).\\
\indent From formulas (1.1)-(1.3), we have
\begin{equation*}
\begin{split}
&\varrho_{g_{\alpha}}(T(X))+\lambda_{1}( \varrho_{g_{\alpha}} ({T_{I_{h}}(X)})-L_{1})+\lambda_{2}(\varrho_{g_{\alpha}} ({T_{R_{h}}(X)})- L_{2})+\lambda_{3}(\Pi_{h} (X)- L_{3})
\\=&m_{1}\varrho_{g_{\alpha}}(X) - m_{2}\varrho_{g_{\alpha}}(h(X)) + m_{3} \varrho_{g_{\gamma}}(h(X)) - D,
\end{split}
\end{equation*}
so (2.10) is expressed as
\begin{equation}
\begin{split}
\min L^{3}_{h}(X)= \underset{h\in\cal{H}}\min\{m_{1}\varrho_{g_{\alpha}}(X)- m_{2}\varrho_{g_{\alpha}}(h(X)) + m_{3} \varrho_{g_{\gamma}}(h(X)) - D \}.
\end{split}
\end{equation}
With the expression (2.5), we have
\begin{equation}
\begin{split}
L^{3}_{f^{*}}(X)=\min L^{3}_{h}(X).
\end{split}
\end{equation}
From Cai et al. (2008) with formulas (2.5) and (2.6), we obtain the following lemma.
\begin{lemma}  For any $h(x)=\sum_{j=1}^nC_{n,j}(x-d_{n,j})_{+}\in \cal{H}$ and given the confidence levels $1-\alpha$ with $0<\alpha<S_{X}(0)$ and $1-\gamma$ with $0<\gamma<S_{X}(0)$, we obtain
\begin{equation*}
\begin{split}
L^{3}_{h}(X)=&m_{1}\int^{1}_{0}  S^{-1}_{X}(t)dg_{\alpha}(t) - m_{2}\Bigg[ \sum_{i=1}^{n-1} \sum_{j=1} ^{i}C_{n,j}\int ^{S_{X}(d_{n,i})}_{S_{X}(d_{n,i+1})}(S^{-1}_{X}(t)-d_{n,j} )dg_{\alpha}(t)
\\&+\sum_{j=1}^{n} C_{n,j}\int ^{S_{X}(d_{n,n})}_{0}(S^{-1}_{X}(t)-d_{n,j} )dg_{\alpha}(t)\Bigg]+m_{3}\Bigg[\sum_{i=1}^{n-1} \sum_{j=1} ^{i}C_{n,j}\int ^{S_{X}(d_{n,i})}_{S_{X}(d_{n,i+1})}(S^{-1}_{X}(t)-d_{n,j} )dg_{\gamma}(t)
\\&+\sum_{j=1}^{n} C_{n,j}\int ^{S_{X}(d_{n,n})}_{0}(S^{-1}_{X}(t)-d_{n,j} )dg_{\gamma}(t)\Bigg ]-D.
\end{split}
\end{equation*}
\end{lemma}
Based on the expression of $L^{3}_{h}(X)$, we analyze its minimum by discussing the magnitude not only of $ m_{2} $ and 0, but also of $M$, $\overline{K}$ and $\underline{K}$. The results are summarized in the following lemma.
\begin{lemma} Given the confidence levels $1-\alpha$ with $0<\alpha<S_{X}(0)$ and $1-\gamma$ with $0<\gamma<S_{X}(0)$, for any function $h(x)=\sum_{j=1}^nC_{n,j}(x-d_{n,j})_{+}\in \cal{H}$ with given coefficients $C_{n,j},$ $j=1,2,\ldots,n$:\\
(1) If $m_{2} = 0$, then $h(x)=0.$\\
(2) When $m_{2} \neq 0$, considering the following cases.\\
(i) If $M \leq\underline{K},$ then
\begin{equation*}
h(x)=x \mathbb{I}_{\{m_{2}>0\}}\sum_{j=1}^{n}C_{n,j}.
\end{equation*}
(ii) If $M \geq\overline{K}$, then
\begin{equation*}
h(x)=x \mathbb{I}_{\{m_{2}<0\}}\sum_{j=1}^{n}C_{n,j}.
\end{equation*}
(iii) When $\underline{K}<M<\overline{K}$, considering the following six cases.\\
Case A. If $0< S_{X}(d_{n,n})\leq S_{X}(d_{n,n-1})\leq \ldots \leq S_{X}(d_{n,1})\leq \widehat{a}$, then
\begin{equation*}
h(x)=(x-S^{-1}_{X}(\widehat{a}))_{+}\mathbb{I}_{\{m_{2}<0\}}\sum_{j=1}^{n}C_{n,j}.
\end{equation*}
Case B. If $\widehat{a}\leq S_{X}(d_{n,n})\leq S_{X}(d_{n,n-1})\leq \ldots \leq S_{X}(d_{n,1})\leq \widehat{b} $, then
\begin{equation*}
h(x)=(x-S^{-1}_{X}(\widehat{b}))_{+}\mathbb{I}_{\{m_{2}>0\}}\sum_{j=1}^{n}C_{n,j}+(x-S^{-1}_{X}(\widehat{a}))_{+}\mathbb{I}_{\{m_{2}<0\}}\sum_{j=1}^{n}C_{n,j}.
\end{equation*}
Case C. If $\widehat{b}\leq S_{X}(d_{n,n})\leq S_{X}(d_{n,n-1})\leq \ldots \leq S_{X}(d_{n,1})<1$, then
\begin{equation*}
h(x)=(x-S^{-1}_{X}(\widehat{b}))_{+}\mathbb{I}_{\{m_{2}>0\}}\sum_{j=1}^{n}C_{n,j}+x\mathbb{I}_{\{m_{2}<0\}}\sum_{j=1}^{n}C_{n,j}.
\end{equation*}
Case D. If $0<S_{X}(d_{n,n})\leq\ldots\leq S_{X}(d_{n,k})\leq\widehat{a}\leq S_{X}(d_{n,k-1}) \leq\ldots\leq S_{X}(d_{n,1})\leq \widehat{b}$, where $k=2,3,\ldots, n,$ then
\begin{equation*}
h(x)=(x-S^{-1}_{X}(\widehat{b}))_{+}\mathbb{I}_{\{m_{2}>0\}}\sum_{j=1}^{k-1}C_{n,j}+(x-S^{-1}_{X}(\widehat{a}))_{+}\mathbb{I}_{\{m_{2}<0\}}\Bigg[\sum_{j=1}^{k-1}C_{n,j}+ \sum_{j=k}^{n}C_{n,j}\Bigg].
\end{equation*}
Case E. If $\widehat{a}\leq S_{X}(d_{n,n})\leq\ldots\leq S_{X}(d_{n,l})\leq\widehat{b}\leq S_{X}(d_{n,l-1}) \leq\ldots\leq S_{X}(d_{n,1})<1$, where $l=2,3,\ldots, n,$ then
\begin{equation*}
h(x)=(x-S^{-1}_{X}(\widehat{b}))_{+}\mathbb{I}_{\{m_{2}>0\}}\Bigg[\sum_{j=1}^{l-1}C_{n,j}+\sum_{j=l}^{n}C_{n,j}\Bigg]
+\mathbb{I}_{\{m_{2}<0\}}\Bigg[x\sum_{j=1}^{l-1}C_{n,j}+(x-S^{-1}_{X}(\widehat{a}))_{+}\sum_{j=l}^{n}C_{n,j}\Bigg].
\end{equation*}
Case F. If $0< S_{X}(d_{n,n})\leq \ldots \leq S_{X}(d_{n,k})\leq \widehat{a} \leq S_{X}(d_{n,k-1}) \leq \ldots \leq S_{X}(d_{n,l})\leq \widehat{b}\leq S_{X}(d_{n,l-1})\leq\ldots \leq S_{X}(d_{n,1})<1$, where $l=2,3,\ldots,k-1,$ $k=3,4,\ldots,n,$ then
\begin{equation*}
\begin{split}
h(x)=&(x-S^{-1}_{X}(\widehat{b}))_{+}\mathbb{I}_{\{m_{2}>0\}}\Bigg[ \sum_{j=1}^{l-1}C_{n,j}+\sum_{j=l}^{k-1}C_{n,j}\Bigg]
\\&+\mathbb{I}_{\{m_{2}<0\}}\Bigg[x\sum_{j=1}^{l-1}C_{n,j}
+(x-S^{-1}_{X}(\widehat{a}))_{+}\Bigg(\sum_{j=l}^{k-1}C_{n,j}+\sum_{j=k}^{n}C_{n,j}\Bigg)
\Bigg].
\end{split}
\end{equation*}
\end{lemma}
\noindent{\bfseries Proof.} By Lemma 3.1, we have
\begin{equation*}
\begin{split}
\frac{\partial L^{3}_{h}(X)}{\partial d_{n,1}}=C_{n,1}[m_{2}g_{\alpha} (S_{X}(d_{n,1}) ) -m_{3}g_{\gamma} (S_{X}(d_{n,1}) )  ],
\\  \frac{\partial L^{3}_{h}(X)}{\partial d_{n,2}}=C_{n,2}[m_{2}g_{\alpha} (S_{X}(d_{n,2}) ) -m_{3}g_{\gamma} (S_{X}(d_{n,2}) )  ],
\end{split}
\end{equation*}
\begin{equation*}
\begin{split}
\vdots
\end{split}
\end{equation*}
\begin{equation*}
\begin{split}
\frac{\partial L^{3}_{h}(X)}{\partial d_{n,n}}=C_{n,n}[m_{2}g_{\alpha} (S_{X}(d_{n,n}) ) -m_{3}g_{\gamma} (S_{X}(d_{n,n}) )  ].
\end{split}
\end{equation*}
\indent With the expression (2.15), if $$\frac {\partial L^{3}_{h}(X)}{\partial d_{n,j}}=0,$$ then $H(d_{n,j})=0.$ Let ${t = S_{X}(x) }$, from (2.14) we know $\underline{K}\leq K(t)\leq\overline{K}$ for any $t\in(0,1)$.
By the definitions of $K(t)$, $M$ and $H(x)$, we derive four cases: if $m_{2} > 0$ and $K(t) \geq M$, then $L^{3}_{h}(X)$ is increasing;
if $m_{2} > 0$ and $K(t) \leq M$, then $L^{3}_{h}(X)$ is decreasing;
if $m_{2} < 0$ and $K(t)\geq M$, then $L^{3}_{h}(X)$ is decreasing;
if $m_{2} < 0$ and $K(t)\leq M$, then $L^{3}_{h}(X)$ is increasing.
In the next we will consider the following possible situations depending on above four cases.\\
\indent 1. If $m_{2} = 0$, then $m_{3}\geq0$, $L^{3}_{h}(X)$ is decreasing. Thus,
the minimum of $ L^{3}_{h}(X)$ is attained at $d_{n,1}=d_{n,2}=\ldots =d_{n,n}=\overline{X},$
$L^{3}_{h}(X)= m_{1}\varrho_{g_{\alpha}}(X)-D,$ $h(x)=0.$\\
\indent 2. When $m_{2}\neq 0$, we consider the following three cases: $M \leq \underline{K}$, $M \geq \overline{K}$ and $\underline{K}<M<\overline{K}$.\\
\indent (1) When $M \leq \underline{K}$, for any $t\in(0,1)$, $K(t)\geq M,$\\
\indent a) if $m_{2}>0,$ then $L^{3}_{h}(X)$ is increasing, and the minimum of $ L^{3}_{h}(X)$ is attained at $d_{n,1}=d_{n,2}=\ldots =d_{n,n}=0,$ so
$$h(x)=\sum_{j=1}^{n}C_{n,j}x,$$ $$L^{3}_{h}(X)= m_{1}\varrho_{g_{\alpha}}(X)-\sum_{j=1}^{n}C_{n,j}\Bigg[\int_{0}^{\infty} H(x)dx\Bigg] -D.$$\\
\indent b) if $m_{2}<0,$ then $L^{3}_{h}(X)$ is decreasing, and the minimum of $ L^{3}_{h}(X)$ is attained at $d_{n,1}=d_{n,2}=\ldots =d_{n,n}=\overline{X},$ so
$ L^{3}_{h}(X)= m_{1}\varrho_{g_{\alpha}}(X)-D,$ $h(x)=0.$\\
\indent (2) When $M \geq \overline{K}$, for any $t\in(0,1)$, $K(t)\leq M,$\\
\indent a) if $m_{2}>0$, then $L^{3}_{h}(X)$ is decreasing and the minimum $ L^{3}_{h}(X)$ is attained at $d_{n,1}=d_{n,2}=\ldots =d_{n,n}=\overline{X},$ so $ L^{3}_{h}(X)= m_{1}\varrho_{g_{\alpha}}(X)-D$, $h(x)=0.$\\
\indent b) if $m_{2}<0$, then $L^{3}_{h}(X)$ is increasing and the minimum $ L^{3}_{h}(X)$ is attained at $d_{n,1}=d_{n,2}=\ldots =d_{n,n}=0 ,$ so $$h(x)=\sum_{j=1}^{n}C_{n,j}x,$$ $$L^{3}_{h}(X)= m_{1}\varrho_{g_{\alpha}}(X)-\sum_{j=1}^{n}C_{n,j}\Bigg[\int_{0}^{\infty} H(x)dx\Bigg] -D.$$\\
\indent (3) When $\underline{K}<M<\overline{K}$, to denote $\widehat{a}=\min\{t:K(t)\geq M\}$ and $\widehat{b}=\max\{t:K(t)\geq M\}$ for $t\in(0,1).$
We obtain that $K(t)\leq M $ on $(0,\widehat{a}]$ and $[\widehat{b}, 1)$, $K(t)\geq M$ on $[\widehat{a}, \widehat{b}]$. If $m_{2}>0$, then $L^{3}_{h}(X)$ is decreasing on $(0,\widehat{a}]$ and $[\widehat{b}, 1)$, and increasing on $[\widehat{a}, \widehat{b}];$ if $m_{2}<0$, then $L^{3}_{h}(X)$ is increasing on $(0,\widehat{a}]$ and $[\widehat{b}, 1)$, and decreasing on $[\widehat{a}, \widehat{b}].$
In the next, we consider the following six cases depending on the relative locations of $S_{X}(d_{n,j})$, $\widehat{a}$ and $ \widehat{b}$.\\
\indent Case A: $0< S_{X}(d_{n,n})\leq S_{X}(d_{n,n-1})\leq \ldots \leq S_{X}(d_{n,1})\leq \widehat{a} $, which is equivalent to $d_{n,n}\geq d_{n,n-1}\geq\ldots \geq d_{n,1} \geq S^{-1}_{X}(\widehat{a})$, in this case $ K(t)\leq M $,\\
\indent a) if $m_{2}>0$, then $L^{3}_{h}(X)$ is decreasing, and the minimum $ L^{3}_{h}(X)$ is attained at $d_{n,1}=d_{n,2}=\ldots =d_{n,n}=\overline{X},$ so $h(x)=0,$ $ L^{3}_{h}(X)= m_{1}\varrho_{g_{\alpha}}(X)-D.$\\
\indent b) if $m_{2}<0$, then $L^{3}_{h}(X)$ is increasing, and the minimum $ L^{3}_{h}(X)$ is attained at $d_{n,1}=d_{n,2}=\ldots =d_{n,n}=S^{-1}_{X}(\widehat{a}),$ so
$$h(x)=\sum_{j=1}^{n}C_{n,j}(x-S^{-1}_{X}(\widehat{a}))_{+},$$
$$ L^{3}_{h}(X)= m_{1}\varrho_{g_{\alpha}}(X)-\sum_{j=1}^{n}C_{n,j}\Bigg[\int_{S^{-1}_{X}(\widehat{a})}^{\infty} H(x)dx\Bigg] -D.$$\\
\indent Case B: $\widehat{a}\leq S_{X}(d_{n,n})\leq S_{X}(d_{n,n-1})\leq \ldots \leq S_{X}(d_{n,1})\leq \widehat{b} $, which is equivalent to $ S^{-1}_{X}(\widehat{a}) \geq d_{n,n}\geq d_{n,n-1}\geq\ldots \geq d_{n,1} \geq S^{-1}_{X}(\widehat{b})$, in this case $ K(t)\geq M $,\\
\indent a) if $m_{2}>0$, then $L^{3}_{h}(X)$ is increasing, and the minimum $ L^{3}_{h}(X)$ is attained at $d_{n,1}=d_{n,2}=\ldots =d_{n,n}=S^{-1}_{X}(\widehat{b}),$ so $$h(x)=\sum_{j=1}^{n}C_{n,j}(x-S^{-1}_{X}(\widehat{b}))_{+},$$
$$L^{3}_{h}(X)= m_{1}\varrho_{g_{\alpha}}(X)-\sum_{j=1}^{n}C_{n,j}\Bigg[\int_{S^{-1}_{X}(\widehat{b})}^{\infty} H(x)dx\Bigg] -D.$$\\
\indent b)  if $m_{2}<0$, then $L^{3}_{h}(X)$ is decreasing, and the minimum $ L^{3}_{h}(X)$ is attained at $d_{n,1}=d_{n,2}=\ldots =d_{n,n}=S^{-1}_{X}(\widehat{a}),$ so $$h(x)=\sum_{j=1}^{n}C_{n,j}(x-S^{-1}_{X}(\widehat{a}))_{+},$$
$$L^{3}_{h}(X)= m_{1}\varrho_{g_{\alpha}}(X)-\sum_{j=1}^{n}C_{n,j}\Bigg[\int_{S^{-1}_{X}(\widehat{a})}^{\infty} H(x)dx\Bigg] -D.$$\\
\indent Case C: $\widehat{b}\leq S_{X}(d_{n,n})\leq S_{X}(d_{n,n-1})\leq \ldots \leq S_{X}(d_{n,1})<1$, which is equivalent to $S^{-1}_{X}(\widehat{b}) \geq d_{n,n}\geq d_{n,n-1}\geq\ldots \geq d_{n,1} $, in this case $ K(x)\leq M $,\\
\indent a) if $m_{2}>0$, then $L^{3}_{h}(X)$ is decreasing, and the minimum $ L^{3}_{h}(X)$ is attained at $d_{n,1}=d_{n,2}=\ldots =d_{n,n}=S^{-1}_{X}(\widehat{b}),$ so
$$h(x)=\sum_{j=1}^{n}C_{n,j}(x-S^{-1}_{X}(\widehat{b}))_{+},$$ $$L^{3}_{h}(X)= m_{1}\varrho_{g_{\alpha}}(X)-\sum_{j=1}^{n}C_{n,j}\Bigg[\int_{S^{-1}_{X}(\widehat{b})}^{\infty} H(x)dx\Bigg] -D.$$\\
\indent b) if $m_{2}<0$, then $L^{3}_{h}(X)$ is increasing, and the minimum $ L^{3}_{h}(X)$ is attained at $d_{n,1}=d_{n,2}=\ldots =d_{n,n}=0,$ so
$$h(x)=\sum_{j=1}^{n}C_{n,j}x,$$ $$ L^{3}_{h}(X)= m_{1}\varrho_{g_{\alpha}}(X)-\sum_{j=1}^{n}C_{n,j}\Bigg[\int_{0}^{\infty} H(x)dx\Bigg] -D. $$\\
\indent Case D: $0<S_{X}(d_{n,n})\leq\ldots\leq S_{X}(d_{n,k})\leq\widehat{a}\leq S_{X}(d_{n,k-1}) \leq\ldots\leq S_{X}(d_{n,1})\leq \widehat{b} $, which is equivalent to $d_{n,n}\geq \ldots \geq d_{n,k}\geq S^{-1}_{X}(\widehat{a})\geq d_{n,k-1}\geq\ldots \geq d_{n,1}\geq S^{-1}_{X}(\widehat{b}) $, where $ k=2,3,\ldots, n, $\\
\indent a) if $m_{2}>0$, then $L^{3}_{h}(X)$ is decreasing on $(0,\widehat{a}]$ and increasing on $[\widehat{a},\widehat{b}]$, the minimum $ L^{3}_{h}(X)$ is attained at $d_{n,1}=\ldots=d_{n,k-1}=S^{-1}_{X}(\widehat{b})$ and $d_{n,k}=\ldots =d_{n,n}=\overline{X},$ so
$$h(x)=\sum_{j=1}^{k-1}C_{n,j}(x-S^{-1}_{X}(\widehat{b}))_{+},$$
$$L^{3}_{h}(X)= m_{1}\varrho_{g_{\alpha}}(X)-\sum_{j=1}^{k-1}C_{n,j}\Bigg[\int_{S^{-1}_{X}(\widehat{b})}^{\infty} H(x)dx\Bigg]-D.$$\\
\indent b) if $m_{2}<0$, then $L^{3}_{h}(X)$ is increasing on $(0,\widehat{a}]$ and decreasing on $[\widehat{a},\widehat{b}]$, the minimum $ L^{3}_{h}(X)$ is attained at $d_{n,1}=\ldots=d_{n,k-1}=S^{-1}_{X}(\widehat{a})$ and $d_{n,k}=\ldots =d_{n,n}=S^{-1}_{X}(\widehat{a})$, so
$$h(x)=\sum_{j=1}^{k-1}C_{n,j}(x-S^{-1}_{X}(\widehat{a}))_{+}+\sum_{j=k}^{n}C_{n,j}(x-S^{-1}_{X}(\widehat{a}))_{+},$$
$$ L^{3}_{h}(X)= m_{1}\varrho_{g_{\alpha}}(X)-\sum_{j=1}^{k-1}C_{n,j}\Bigg[\int_{S^{-1}_{X}(\widehat{a})}^{\infty} H(x)dx\Bigg]-\sum_{j=k}^{n}C_{n,j}\Bigg[\int_{S^{-1}_{X}(\widehat{a})}^{\infty} H(x)dx\Bigg] -D.$$
\indent Case E: $\widehat{a}\leq S_{X}(d_{n,n})\leq\ldots\leq S_{X}(d_{n,l})\leq\widehat{b}\leq S_{X}(d_{n,l-1}) \leq\ldots\leq S_{X}(d_{n,1})<1$, which is equivalent to
$S^{-1}_{X}(\widehat{a})\geq d_{n,n}\geq \ldots \geq d_{n,l}\geq S^{-1}_{X}(\widehat{b})\geq d_{n,l-1}\geq\ldots \geq d_{n,1},$ where $ l=2,3,\ldots, n, $\\
\indent a) if $m_{2}>0$, then $L^{3}_{h}(X)$ is increasing on $[\widehat{a},\widehat{b}]$ and decreasing on $[\widehat{b},1)$, the minimum $ L^{3}_{h}(X)$ is attained at $d_{n,1}=\ldots=d_{n,l-1}=S^{-1}_{X}(\widehat{b})$ and $d_{n,l}=\ldots =d_{n,n}=S^{-1}_{X}(\widehat{b})$, so
$$h(x)=\sum_{j=1}^{l-1}C_{n,j}(x-S^{-1}_{X}(\widehat{b}))_{+}+\sum_{j=l}^{n}C_{n,j}(x-S^{-1}_{X}(\widehat{b}))_{+},$$
$$ L^{3}_{h}(X)= m_{1}\varrho_{g_{\alpha}}(X)-\sum_{j=1}^{l-1}C_{n,j}\Bigg[\int_{S^{-1}_{X}(\widehat{b})}^{\infty} H(x)dx\Bigg]-\sum_{j=l}^{n}C_{n,j}\Bigg[\int_{S^{-1}_{X}(\widehat{b})}^{\infty} H(x)dx\Bigg] -D.$$
\indent b) if $m_{2}<0$, then $L^{3}_{h}(X)$ is decreasing on $[\widehat{a},\widehat{b}]$ and increasing on $[\widehat{b},1)$, the minimum $ L^{3}_{h}(X)$ is attained at $d_{n,1}=\ldots=d_{n,l-1}=0$ and $d_{n,l}=\ldots =d_{n,n}=S^{-1}_{X}(\widehat{a})$, so
$$h(x)=\sum_{j=1}^{l-1}C_{n,j}x+\sum_{j=l}^{n}C_{n,j}(x-S^{-1}_{X}(\widehat{a}))_{+},$$
$$ L^{3}_{h}(X)= m_{1}\varrho_{g_{\alpha}}(X)-\sum_{j=1}^{l-1}C_{n,j}\Bigg[\int_{0}^{\infty} H(x)dx\Bigg]-\sum_{j=l}^{n}C_{n,j}\Bigg[\int_{S^{-1}_{X}(\widehat{a})}^{\infty} H(x)dx\Bigg]-D.$$
\indent Case F: $0< S_{X}(d_{n,n})\leq \ldots \leq S_{X}(d_{n,k})\leq \widehat{a} \leq S_{X}(d_{n,k-1}) \leq \ldots \leq S_{X}(d_{n,l})\leq \widehat{b}\leq S_{X}(d_{n,l-1})\leq\ldots \leq S_{X}(d_{n,1})<1$, which is equivalent to $ d_{n,n}\geq \ldots \geq d_{n,k}\geq S^{-1}_{X}(\widehat{a})  \geq d_{n,k-1}\geq \ldots\geq d_{n,l}\geq S^{-1}_{X}(\widehat{b})  \geq d_{n,l-1} \geq \ldots \geq d_{n,1}\geq 0 $, where $ l=2,3,\ldots,k-1,$  $k=3,4,\ldots,n,$\\
\indent a) if $m_{2}>0$, then $L^{3}_{h}(X)$ is decreasing on $(0,\widehat{a}]$ and $[\widehat{b},1)$, and increasing on $[\widehat{a},\widehat{b}]$, the minimum $ L^{3}_{h}(X)$ is attained at $d_{n,1}=\ldots=d_{n,l-1}=S^{-1}_{X}(\widehat{b})$, $d_{n,l}=\ldots =d_{n,k-1}=S^{-1}_{X}(\widehat{b})$ and $d_{n,k}=\ldots =d_{n,n}=\overline{X}$, so $$h(x)=\sum_{j=1}^{l-1}C_{n,j}(x-S^{-1}_{X}(\widehat{b}))_{+}+\sum_{j=l}^{k-1}C_{n,j}(x-S^{-1}_{X}(\widehat{b}))_{+},$$
$$ L^{3}_{h}(X)= m_{1}\varrho_{g_{\alpha}}(X)-\sum_{j=1}^{l-1}C_{n,j}\Bigg[\int_{S^{-1}_{X}(\widehat{b})}^{\infty} H(x)dx\Bigg]-\sum_{j=l}^{k-1}C_{n,j}\Bigg[\int_{S^{-1}_{X}(\widehat{b})}^{\infty} H(x)dx\Bigg]-D.$$
\indent b) if $m_{2}<0$, then $L^{3}_{h}(X)$ is increasing on $(0,\widehat{a}]$ and $[\widehat{b},1)$, and decreasing on $[\widehat{a},\widehat{b}]$, the minimum $ L^{3}_{h}(X)$ is attained at $d_{n,1}=\ldots=d_{n,l-1}=0$ and $d_{n,l}=\ldots =d_{n,k-1}=S^{-1}_{X}(\widehat{a}),d_{n,k}=\ldots =d_{n,n}=S^{-1}_{X}(\widehat{a})$, so
$$h(x)=\sum_{j=1}^{l-1}C_{n,j}x+\sum_{j=l}^{k-1}C_{n,j}(x-S^{-1}_{X}(\widehat{a}))_{+}+\sum_{j=k}^{n}C_{n,j}(x-S^{-1}_{X}(\widehat{a}))_{+},$$
$$L^{3}_{h}(X)= m_{1}\varrho_{g_{\alpha}}(X)-\sum_{j=1}^{l-1}C_{n,j}\Bigg[\int_{0}^{\infty} H(x)dx\Bigg]-\Bigg(\sum_{j=l}^{k-1}C_{n,j}+\sum_{j=k}^{n}C_{n,j}\Bigg)\Bigg[\int_{S^{-1}_{X}(\widehat{a})}^{\infty} H(x)dx\Bigg]-D.$$ $\hfill\square$\\
\indent We are ready to present the key results of this section which are stated in Theorem 3.1. Lemma 3.2 is used to obtain the solution of (3.1) by determining the value of $\sum C_{ n, j}$. The specific results are summarized in the following theorem.
\begin{theorem} Given a confidence levels $1-\alpha$ with $0<\alpha<S_{X}(0)$ and $1-\gamma$ with $0<\gamma<S_{X}(0)$, for any function $h(x)=\sum_{j=1}^nC_{n,j}(x-d_{n,j})_{+}\in \cal{H}$ with given coefficients $C_{n,j},$ $j=1,2,\ldots,n$:\\
(1) If $m_{2}=0$, then $f^{*}(x)=0$.\\
(2) When $m_{2}\neq0$, considering the following cases.\\
(i) If $M<\underline{K}$, then
\begin{equation*}
f^{*}(x)=x\mathbb{I}_{\{m_{2}>0\}}.
\end{equation*}
(ii) When $M=\underline{K}$, if there exist a point $t_{0}\in(0,1)$ such that $K(t_{0})=M$, then $f^{*}\in\cal{H}$; for other cases,
\begin{equation*}
f^{*}(x)=x\mathbb{I}_{\{m_{2}>0\}}.
\end{equation*}
(iii) If $M>\overline{K}$, then
\begin{equation*}
f^{*}(x)=x\mathbb{I}_{\{m_{2}<0\}}.
\end{equation*}
(iv) When $M=\overline{K}$, if there exist a point $t^{*}\in(0,1)$ such that $K(t^{*})=M$, then $f^{*}\in\cal{H}$; for other cases,
\begin{equation*}
f^{*}(x)=x\mathbb{I}_{\{m_{2}<0\}}.
\end{equation*}
(v) When $\underline{K}<M<\overline{K}$, we consider the following six cases.\\
Case A. If $t=\widehat{a}$, then $f^{*}\in\cal{H}$. For $0< S_{X}(d_{n,n})\leq S_{X}(d_{n,n-1})\leq \ldots \leq S_{X}(d_{n,1})< \widehat{a}$,
\begin{equation*}
f^{*}(x)=(x-S^{-1}_{X}(\widehat{a}))_{+}\mathbb{I}_{\{m_{2}<0\}}.
\end{equation*}
Case B. If $t=\widehat{a}$ or $t=\widehat{b}$, then $f^{*}\in\cal{H}$. For $\widehat{a}< S_{X}(d_{n,n})\leq S_{X}(d_{n,n-1})\leq \ldots \leq S_{X}(d_{n,1})< \widehat{b}$,
\begin{equation*}
f^{*}(x)=(x-S^{-1}_{X}(\widehat{b}))_{+}\mathbb{I}_{\{m_{2}>0\}}.
\end{equation*}
Case C. If $t=\widehat{b}$, then $f^{*}\in\cal{H}$. For $\widehat{b}< S_{X}(d_{n,n})\leq S_{X}(d_{n,n-1})\leq \ldots \leq S_{X}(d_{n,1})<1$,
\begin{equation*}
f^{*}(x)=x\mathbb{I}_{\{m_{2}<0\}}.
\end{equation*}
Case D. If $t=\widehat{a}$ or $t=\widehat{b}$, then $f^{*}\in\cal{H}$. For $0<S_{X}(d_{n,n})\leq\ldots\leq S_{X}(d_{n,k})<\widehat{a}< S_{X}(d_{n,k-1}) \leq\ldots\leq S_{X}(d_{n,1})< \widehat{b}$, where $k=2,3,\ldots, n,$
\begin{equation*}
f^{*}(x)=(x-S^{-1}_{X}(\widehat{b}))_{+}\mathbb{I}_{\{m_{2}>0\}}+(x-S^{-1}_{X}(\widehat{a}))_{+}\mathbb{I}_{\{m_{2}<0\}}.
\end{equation*}
Case E. If $t=\widehat{a}$ or $t=\widehat{b}$, then $f^{*}\in\cal{H}$. For $\widehat{a}< S_{X}(d_{n,n})\leq\ldots\leq S_{X}(d_{n,l})<\widehat{b}< S_{X}(d_{n,l-1}) \leq\ldots\leq S_{X}(d_{n,1})<1$, where $l=2,3,\ldots, n,$
\begin{equation*}
f^{*}(x)=(x-S^{-1}_{X}(\widehat{b}))_{+}\mathbb{I}_{\{m_{2}>0\}}+x\mathbb{I}_{\{m_{2}<0\}}.
\end{equation*}
Case F. If $t=\widehat{a}$ or $t=\widehat{b}$, then $f^{*}\in\cal{H}$. For $0< S_{X}(d_{n,n})\leq \ldots \leq S_{X}(d_{n,k})< \widehat{a} < S_{X}(d_{n,k-1}) \leq \ldots \leq S_{X}(d_{n,l})< \widehat{b}< S_{X}(d_{n,l-1})\leq\ldots \leq S_{X}(d_{n,1})<1$, where $l=2,3,\ldots,k-1,$ $k=3,4,\ldots,n,$
\begin{equation*}
f^{*}(x)=(x-S^{-1}_{X}(\widehat{b}))_{+}\mathbb{I}_{\{m_{2}>0\}}+[c_{1}x+c_{2}(x-S^{-1}_{X}(\widehat{a}))_{+}]\mathbb{I}_{\{m_{2}<0\}}.
\end{equation*}
where $c_{1}>0,c_{2}>0,$ and $c_{1}x+c_{2}(x-S^{-1}_{X}(\widehat{a}))_{+}\leq x.$
\end{theorem}
\noindent{\bfseries Proof.} To obtain the specific form of the optimal ceded loss function $f ^{*}$, we only need to judge the magnitude of $ \sum C_{ n, j} $ according to the sign of $H(x)$.\\
\indent 1. Since $m_{2}=0$, $L^{3}_{h}(X)= m_{1}\varrho_{g_{\alpha}}(X)-D,$ $h(x)=0.$ From (3.2) we derive that $L^{3}_{f^{*}}(X)= m_{1}\rho_{g_{\alpha}}(X)-D,$ $f^{*}=0.$\\
\indent 2. When $m_{2}\neq0$, we consider the following cases.\\
\indent (1) When $ M<\underline{K}$, for any $ S_{X}(x)\in (0,1),$  $K(S_{X}(x))> M $, which is equivalent to $$\frac{g_{\alpha}(S_{X}(x))}{g_{\gamma}(S_{X}(x))}> \frac{m_{3}}{m_{2}},$$
\indent a) if $m_{2}>0,$ then
$$H(x) =m_{2}g_{\alpha}(S_{X}(x))-m_{3} g_{\gamma}(S_{X}(x))> 0.$$
We derive that when $\sum_{j=1}^{n}C_{n,j}=1$,
$ L^{3}_{f^{*}}(X)= m_{1}\varrho_{g_{\alpha}}(X)-\int_{0}^{\infty} H(x)dx -D,$ the minimum of $ L^{3}_{h}(X)$ is attained at
$f^{*}=x.$\\
\indent b) if $m_{2}<0 ,$ then $ L^{3}_{f^{*}}(X)= m_{1}\varrho_{g_{\alpha}}(X)-D,$
the minimum of $ L^{3}_{h}(X)$ is attained at $f^{*}=0.$\\
\indent (2) When $ M=\underline{K}$, if there exist a point $t_{0}$ such that $K(t_{0})=M$, then $H(x)=0$, $f^{*}\in\cal{H}$.
For other cases, the proof is similar to the case of $ M<\underline{K}$ and we omit it.\\
\indent (3) When $M>\overline{K}$, for any $ S_{X}(x)\in (0,1),$  $K(S_{X}(x))< M $,\\
\indent a) if $m_{2}>0,$ then $ L^{3}_{f^{*}}(X)= m_{1}\varrho_{g_{\alpha}}(X)-D,$ the minimum of $ L^{3}_{h}(X)$ is attained at
$f^{*}=0.$\\
\indent b) if $m_{2}<0 ,$ then $H(x)>0$. We derive that when $\sum_{j=1}^{n}C_{n,j}=1$,
$ L^{3}_{f^{*}}(X)= m_{1}\varrho_{g_{\alpha}}(X)-\int_{0}^{\infty} H(x)dx -D,$ the minimum of $ L^{3}_{h}(X)$ is attained at
$f^{*}=x.$\\
\indent (4) When $M=\overline{K}$, if there exist a point $t^{*}$ such that $K(t^{*})=M$, then $H(x)=0$, $f^{*}\in\cal{H}$.
For other cases, the proof is similar to the case of $M>\overline{K}$ and we omit it.\\
\indent (5) When $\underline{K}< M < \overline{K}$, we consider the following six cases:\\
\indent i) for Case A: if $t=\widehat{a}$, then $K(t)= M$, $H(x)=0$, therefore, $f^{*}\in\cal{H}$.
For any $S_{X}(d_{n,j})\in(0,\widehat{a})$, $K(S_{X}(d_{n,j}))< M$, we consider the following two cases,\\
\indent a) if $m_{2}>0$, then
$ L^{3}_{f^{*}}(X)= m_{1}\varrho_{g_{\alpha}}(X)-D,$
the minimum of $ L^{3}_{h}(X)$ is attained at $f^{*}=0.$\\
\indent b) if $m_{2}<0$, then $H(x) > 0,$ therefore, when $\sum_{j=1}^{n}C_{n,j}=1$,
$$L^{3}_{f^{*}}(X)= m_{1}\varrho_{g_{\alpha}}(X)-\int_{S^{-1}_{X}(\widehat{a})}^{\infty} H(x)dx -D,$$
the minimum of $ L^{3}_{h}(X)$ is attained at $f^{*}=(x-S^{-1}_{X}(\widehat{a}))_{+}.$\\
\indent ii) for Case B: if $t=\widehat{a}$ or $t=\widehat{b}$, then $K(t)= M$, $H(x)=0$, therefore, $f^{*}\in\cal{H}$.
For any $ S_{X}(d_{n,j})\in (\widehat{a},\widehat{b})$, $K(S_{X}(d_{n,j})> M,$ we consider the following two cases,\\
\indent a) if $m_{2}>0$, then $H(x) > 0,$ when $\sum_{j=1}^{n}C_{n,j}=1$,
$$ L^{3}_{f^{*}}(X)= m_{1}\varrho_{g_{\alpha}}(X)-\int_{S^{-1}_{X}(\widehat{b})}^{\infty} H(x)dx -D,$$
the minimum of $ L^{3}_{h}(X)$ is attained at $f^{*}=(x-S^{-1}_{X}(\widehat{b}))_{+}.$\\
\indent b) if $m_{2}<0$, then $H(x) < 0,$ when $\sum_{j=1}^{n}C_{n,j}=0$,
$L^{3}_{f^{*}}(X)= m_{1}\varrho_{g_{\alpha}}(X) -D,$
the minimum of $ L^{3}_{h}(X)$ is attained at $f^{*}=0.$\\
\indent iii) for Case C: if $t=\widehat{b}$, then $K(t)= M$, $H(x)=0$, therefore, $f^{*}\in\cal{H}$.
For any $S_{X}(d_{n,j})\in(\widehat{b},1)$, $K(S_{X}(d_{n,j})< M,$ we consider the following two cases,\\
\indent a) if $m_{2}>0$, then $H(x)< 0,$ when $\sum_{j=1}^{n}C_{n,j}=0 $,
$L^{3}_{f^{*}}(X)= m_{1}\varrho_{g_{\alpha}}(X) -D,$ the minimum of $ L^{3}_{h}(X)$ is attained at $f^{*}=0.$\\
\indent b) if $m_{2}<0$, then $H(x) > 0,$ when $\sum_{j=1}^{n}C_{n,j}=1$,
$$ L^{3}_{f^{*}}(X)= m_{1}\varrho_{g_{\alpha}}(X)- \int_{0}^{\infty} H(x)dx-D, $$
the minimum of $ L^{3}_{h}(X)$ is attained at $f^{*}=x.$\\
\indent iv) for Case D: if $t=\widehat{a}$ or $t=\widehat{b}$, then $K(t)= M$, $H(x)=0$, therefore, $f^{*}\in\cal{H}$.
For other cases, we obtain that $K(S_{X}(d_{n,j}))< M$ for $S_{X}(d_{n,j})\in(0,\widehat{a})$, where $ j=k,k+1,\ldots, n, $
$K(S_{X}(j))> M$ for $S_{X}(d_{n,j})\in(\widehat{a},\widehat{b})$, where $j=1,2,\ldots, k-1,$\\
\indent a) if $m_{2}>0$, then $H(x)< 0$ for $S_{X}(d_{n,j})\in(0,\widehat{a})$, where $ j=k,k+1,\ldots, n;$
$H(x) > 0$ for $S_{X}(d_{n,j})\in(\widehat{a},\widehat{b})$, where $j=1,2,\ldots, k-1.$
So, when $\sum_{j=1}^{k-1}C_{n,j}=1$ and $\sum_{j=k}^{n}C_{n,j}=0$,
$$ L^{3}_{f^{*}}(X)= m_{1}\varrho_{g_{\alpha}}(X)-\int_{S^{-1}_{X}(\widehat{b})}^{\infty}H(x)dx-D,$$
the minimum of $ L^{3}_{h}(X)$ is attained at $f^{*}=(x-S^{-1}_{X}(\widehat{b}))_{+}.$\\
\indent b) if $m_{2}<0$, then $H(x) > 0$ for $S_{X}(d_{n,j})\in(0,\widehat{a})$, where $ j=k,k+1,\ldots, n;$ $H(x)< 0$ for $S_{X}(d_{n,j})\in(\widehat{a},\widehat{b})$, where $j=1,2,\ldots, k-1.$
So, when ${\sum_{j=1}^{k-1}C_{n,j}=0}$ and $\sum_{j=k}^{n}C_{n,j}=1$,
$$ L^{3}_{f^{*}}(X)= m_{1}\varrho_{g_{\alpha}}(X)-\int_{S^{-1}_{X}(\widehat{a})}^{\infty} H(x)dx-D,$$
the minimum of $ L^{3}_{h}(X)$ is attained at $f^{*}=(x-S^{-1}_{X}(\widehat{a}))_{+}.$\\
\indent v) for Case E: if $t=\widehat{a}$ or $t=\widehat{b}$, then $K(t)= M$, $H(x)=0$, therefore, $f^{*}\in\cal{H}$.
For other cases, we obtain that $K(S_{X}(j))> M $ for $S_{X}(d_{n,j})\in(\widehat{a},\widehat{b})$, where $ j=l,l+1,\ldots, n, $
$K(S_{X}(d_{n,j}))< M$ for $S_{X}(d_{n,j})\in(\widehat{b},1)$, where $ j=1,2,\ldots, l-1, $\\
\indent a) if $m_{2}>0$, then  $H(x) >0$ for $S_{X}(d_{n,j})\in(\widehat{a},\widehat{b})$, where $ j=l,l+1,\ldots, n; $
$H(x) < 0$ for $S_{X}(d_{n,j})\in(\widehat{b},1)$, where $ j=1,2,\ldots, l-1.$ We derive that $\sum_{j=1}^{l-1}C_{n,j}=0$ and $\sum_{j=l}^{n}C_{n,j}=1 $,
$$L^{3}_{f^{*}}(X)= m_{1}\varrho_{g_{\alpha}}(X)-\int_{S^{-1}_{X}(\widehat{b})}^{\infty}H(x)dx -D,$$
the minimum of $ L^{3}_{h}(X)$ is attained at $f^{*}=(x-S^{-1}_{X}(\widehat{b}))_{+}.$\\
\indent b) if $m_{2}<0$, then $H(x)<0$ for $S_{X}(d_{n,j})\in(\widehat{a},\widehat{b})$, where $ j=l,l+1,\ldots, n;$
$H(x) > 0$ for $S_{X}(d_{n,j})\in(\widehat{b},1)$, where $ j=1,2,\ldots, l-1.$
We derive that $\sum_{j=1}^{l-1}C_{n,j}=1$ and $\sum_{j=l}^{n}C_{n,j}=0 $,
$$ L^{3}_{f^{*}}(X)= m_{1}\varrho_{g_{\alpha}}(X)-\int_{0}^{\infty} H(x)dx-D,$$
the minimum of $ L^{3}_{h}(X)$ is attained at $f^{*}=x.$\\
\indent vi) for Case F: if $t=\widehat{a}$ or $t=\widehat{b}$, then $K(t)= M$, $H(x)=0$, therefore, $f^{*}\in\cal{H}$.
For other cases, we obtain that $K(S_{X}(d_{n,j}))< M$ for $S_{X}(d_{n,j})\in(0,\widehat{a})$, where $ j=k,k+1,\ldots, n, $
$K(S_{X}(d_{n,j}))> M$ for $S_{X}(d_{n,j})\in(\widehat{a},\widehat{b})$, where $ j=l,l+1,\ldots, k-1, $
$K(S_{X}(d_{n,j}))< M$ for $S_{X}(d_{n,j})\in(\widehat{b},1)$, where $ j=1,2,\ldots, l-1, $\\
\indent a) if $m_{2}>0$, then $H(x)<0$ for $S_{X}(d_{n,j})\in(0,\widehat{a})$, where $ j=k,k+1,\ldots, n;$
$H(x) > 0$ for $S_{X}(d_{n,j})\in(\widehat{a},\widehat{b})$, where $ j=l,l+1,\ldots, k-1;$
$H(x)<0$ for $S_{X}(d_{n,j})\in(\widehat{b},1)$, where $ j=1,2,\ldots, l-1.$
We derive that $\sum_{j=1}^{l-1}C_{n,j}=0$ and $\sum_{j=l}^{k-1}C_{n,j}=1$,
$$ L^{3}_{f^{*}}(X)= m_{1}\varrho_{g_{\alpha}}(X)-\int_{S^{-1}_{X}(\widehat{b})}^{\infty} H(x)dx-D,$$
the minimum of $ L^{3}_{h}(X)$ is attained at $f^{*}=(x-S^{-1}_{X}(\widehat{b}))_{+}.$\\
\indent b) if $m_{2}<0$, then $H(x)>0$ for $S_{X}(d_{n,j})\in(0,\widehat{a})$, where $ j=k,k+1,\ldots, n;$
$H(x) < 0$ for $S_{X}(d_{n,j})\in(\widehat{a},\widehat{b})$, where $ j=l,l+1,\ldots, k-1;$
$H(x)>0$ for $S_{X}(d_{n,j})\in(\widehat{b},1)$, where $ j=1,2,\ldots, l-1.$
We derive that $\sum_{j=1}^{l-1}C_{n,j}=c_{1}$, $\sum_{j=l}^{k-1}C_{n,j}=0$ and $\sum_{j=k}^{n}C_{n,j}=c_{2}$,
$$L^{3}_{f^{*}}(X)= m_{1}\varrho_{g_{\alpha}}(X)-c_{1} \int_{0}^{\infty} H(x)dx-c_{2} \int_{S^{-1}_{X}(\widehat{a})}^{\infty} H(x)dx-D,$$
the minimum of $ L^{3}_{h}(X)$ is attained at
$f^{*}=c_{1}x+c_{2}(x-S^{-1}_{X}(\widehat{a}))_{+},$ where $c_{1}>0, c_{2}>0,$ and $c_{1}x+c_{2}(x-S^{-1}_{X}(\widehat{a}))_{+}\leq x.$ $\hfill\square$
\begin{remark}
\indent When $\lambda_{1}=\lambda_{2}=\lambda_{3}=0$ and $\beta=1$, if we adopt $\rm{VaR}$ risk measure and Wang's premium principle, then our results recover Theorem 3 of Cheung (2010) and superior to them since their results only consist of quota-share reinsurance, but our results consist of the quota-share, stop-loss, change-loss, the combination of quota-share and change-loss or the combination of change-loss and change-loss with different retentions, which means that our results provide more options for reinsurance strategies.
\end{remark}
\begin{remark}
\indent In this paper, we only give the proof of the case that $K(t)$ is a concave function. When $K(t)$ has an irregular shape in the interval (0,1), it is easy to derive the optimal reinsurance with the same way according to Remark 4.3 of Zhuang et al. (2016).\vspace{-1.5em}
\end{remark}
\indent From Theorem 3.1, we obtain the other cases of unconstrained optimal problem and constrained optimal problems. Next, we will only give the unconstrained optimal reinsurance and other cases achieve with the same way.
\begin{corollary}(Unconstrained optimization problem)  Given the confidence levels $1-\alpha$ with $0<\alpha<S_{X}(0)$ and $1-\gamma$ with $0<\gamma<S_{X}(0)$, when $\lambda_{1}=\lambda_{2}=\lambda_{3}=0,$ from Theorem 3.1 we achieve the followings.\\
\indent (1) If $\beta=\frac{1}{2},$ then $f^{*}(x)\in \cal{H}.$\\
\indent (2) When $\beta\neq\frac{1}{2},$ considering the following cases.\\
\indent (i) When $ 1+\rho\leq\underline{K}$, if there exist a point $t_{0}$ such that $K(t_{0})=1+\rho$, then $f^{*}(x)\in \cal{H};$ for other cases,
\begin{equation*}
f^{*}(x)=x\mathbb{I}_{\{\beta>\frac{1}{2}\}}.
\end{equation*}
\indent (ii) When $ 1+\rho\geq\overline{K}$, if there exist a point $t^{*}$ such that $K(t^{*})=1+\rho$, then $f^{*}(x)\in \cal{H};$ for other cases,
\begin{equation*}
f^{*}(x)=x\mathbb{I}_{\{\beta<\frac{1}{2}\}}.
\end{equation*}
\indent (iii) When $\underline{K}<1+\rho<\overline{K}$, considering the following six cases.\\
\indent Case A. If $t=\widehat{a}$, then $f^{*}\in\cal{H}$. For $0< S_{X}(d_{n,n})\leq S_{X}(d_{n,n-1})\leq \ldots \leq S_{X}(d_{n,1})< \widehat{a}$,
\begin{equation*}
f^{*}(x)=(x-S^{-1}_{X}(\widehat{a}))_{+}\mathbb{I}_{\{\beta<\frac{1}{2}\}}.
\end{equation*}
\indent Case B. If $t=\widehat{a}$ or $t=\widehat{b}$, then $f^{*}\in\cal{H}$. For $\widehat{a}< S_{X}(d_{n,n})\leq S_{X}(d_{n,n-1})\leq \ldots \leq S_{X}(d_{n,1})< \widehat{b} $,
\begin{equation*}
f^{*}(x)=(x-S^{-1}_{X}(\widehat{b}))_{+}\mathbb{I}_{\{\beta>\frac{1}{2}\}}.
\end{equation*}
\indent Case C. If $t=\widehat{b}$, then $f^{*}\in\cal{H}$. For $\widehat{b}< S_{X}(d_{n,n})\leq S_{X}(d_{n,n-1})\leq \ldots \leq S_{X}(d_{n,1})<1$,
\begin{equation*}
f^{*}(x)=x\mathbb{I}_{\{\beta<\frac{1}{2}\}}.
\end{equation*}
\indent Case D. If $t=\widehat{a}$ or $t=\widehat{b}$, then $f^{*}\in\cal{H}$. For $0<S_{X}(d_{n,n})\leq\ldots\leq S_{X}(d_{n,k})<\widehat{a}< S_{X}(d_{n,k-1}) \leq\ldots\leq S_{X}(d_{n,1})< \widehat{b}$, where $k=2,3,\ldots, n,$
\begin{equation*}
f^{*}(x)=(x-S^{-1}_{X}(\widehat{b}))_{+}\mathbb{I}_{\{\beta>\frac{1}{2}\}}+(x-S^{-1}_{X}(\widehat{a}))_{+}\mathbb{I}_{\{\beta<\frac{1}{2}\}}.
\end{equation*}
\indent Case E. If $t=\widehat{a}$ or $t=\widehat{b}$, then $f^{*}\in\cal{H}$. For $\widehat{a}< S_{X}(d_{n,n})\leq\ldots\leq S_{X}(d_{n,l})<\widehat{b}< S_{X}(d_{n,l-1}) \leq\ldots\leq S_{X}(d_{n,1})<1$, where $l=2,3,\ldots, n,$
\begin{equation*}
f^{*}(x)=(x-S^{-1}_{X}(\widehat{b}))_{+}\mathbb{I}_{\{\beta>\frac{1}{2}\}}+ x\mathbb{I}_{\{\beta<\frac{1}{2}\}}.
\end{equation*}
\indent Case F. If $t=\widehat{a}$ or $t=\widehat{b}$, then $f^{*}\in\cal{H}$. For $0< S_{X}(d_{n,n})\leq \ldots \leq S_{X}(d_{n,k})< \widehat{a} < S_{X}(d_{n,k-1}) \leq \ldots \leq S_{X}(d_{n,l})< \widehat{b}< S_{X}(d_{n,l-1})\leq\ldots \leq S_{X}(d_{n,1})<1$, where $l=2,3,\ldots,k-1,$ $k=3,4,\ldots,n,$
\begin{equation*}
f^{*}(x)=(x-S^{-1}_{X}(\widehat{b}))_{+}\mathbb{I}_{\{\beta>\frac{1}{2}\}}+[c_{1}x+c_{2}(x-S^{-1}_{X}(\widehat{a}))_{+}]\mathbb{I}_{\{\beta<\frac{1}{2}\}}.
\end{equation*}
where $c_{1}>0,c_{2}>0,$ and $c_{1}x+c_{2}(x-S^{-1}_{X}(\widehat{a}))_{+}\leq x.$
\end{corollary}
\begin{remark}
If there no constraints, then $\lambda_{1}=\lambda_{2}=\lambda_{3}=0.$ We know that $M=1+\rho, H(d_{n,j})=(2\beta-1)[g_{\alpha}(S_{X}(d_{n,j}))-(1+\rho)g_{\gamma}(S_{X}(d_{n,j}))]$. If $\beta=\frac{1}{2},$ $t=t_{0}$, $t=t^{*}$, $t=\widehat{a}$ or $t=\widehat{b}$, then $f^{*}\in \cal{H},$ which means that optimal reinsurance strategy could be any increasing convex function.
\end{remark}
 \section{Two special cases}
\setcounter{equation}{0}
\indent{In this section, we consider two special cases: Value at Risk (VaR) and Tail Value at Risk (TVaR). In order to simple calculation, we take the expectation premium principle, which means that the risk of the insurer and the reinsurer are measured by VaR and TVaR risk measures under the expectation premium principle. Next, we will give the optimal reinsurance under the VaR risk measure and a corresponding numerical example.}
\vskip 0.2cm
 \subsection{Value at Risk}
\indent As we all know, the Value at Risk is a special example of the distortion risk measure with the distortion function
\begin{equation*}
g_{\alpha}(t)=\left\{ \begin{aligned}   \overset{} 0, \qquad  t<\alpha, \\ 1 ,  \qquad t\geq \alpha. \end{aligned}   \right.
\end{equation*}
\noindent When adopting the expectation premium principle $g_{\gamma}(t)=t$, we derive
\begin{equation}
K(t)= \frac{g_{\alpha}(t)}{g_{\gamma}(t)}=\left\{ \begin{aligned}   \overset{} 0, \qquad  t<\alpha, \\ \frac{1}{t} ,  \qquad t\geq \alpha. \end{aligned}   \right.
\end{equation}
\vspace{3ex}From Theorem 3.1, we obtain the following proposition.\\
{\bfseries Proposition 4.1.} Assuming the risk is measured by the VaR risk measure under the expectation premium principle, for $0<\alpha<S_{X}(0)$ and  $0<\alpha<S_{X}(0)$, $\beta\in[0,1]$, $\lambda_{i}\geq0$, where $i=1,2,3,$ we can derive the following results.\\
\indent (1) If $m_{2}=0$, then $f^{*}(x)=0$.\\
\indent (2) When $m_{2}\neq0$, considering the following cases.\\
\indent (i) When $M\leq0$, if $t\in(0,\alpha)$ and $M=0$, then $K(t)=M$, $f^{*}(x)\in \cal{H};$ for other cases, $f^{*}(x)=0.$\\
\indent (ii) When $M\geq\frac{1}{\alpha}$, if $t^{*}=\alpha$ and $M=\frac{1}{\alpha}$, then $K(t^{*})=M$, $f^{*}(x)\in \cal{H};$ for other cases,
\begin{equation*}
f^{*}(x)=x\mathbb{I}_{\{m_{2}<0}\}.
\end{equation*}
\indent (iii) When $0<M<\frac{1}{\alpha}$, we consider the following six cases.\\
\indent Case A. If $t=\alpha$, then $f^{*}\in\cal{H}$. For $0< S_{X}(d_{n,n})\leq S_{X}(d_{n,n-1})\leq \ldots \leq S_{X}(d_{n,1})< \alpha$,
\begin{equation*}
f^{*}(x)=(x-{\rm VaR}_{\alpha}(X))_{+}\mathbb{I}_{\{m_{2}<0}\}.
\end{equation*}
\indent Case B. If $t=\alpha$ or $t=\widehat{b}$, then $f^{*}\in\cal{H}$. For $\alpha< S_{X}(d_{n,n})\leq S_{X}(d_{n,n-1})\leq \ldots \leq S_{X}(d_{n,1})< \widehat{b} $,  \begin{equation*}
f^{*}(x)=(x-S^{-1}_{X}(\widehat{b}))_{+}\mathbb{I}_{\{m_{2}>0}\}.
\end{equation*}
\indent Case C. If $t=\widehat{b}$, then $f^{*}\in\cal{H}$. For $\widehat{b}< S_{X}(d_{n,n})\leq S_{X}(d_{n,n-1})\leq \ldots \leq S_{X}(d_{n,1})<1$,
\begin{equation*}
f^{*}(x)= x\mathbb{I}_{\{m_{2}<0}\}.
\end{equation*}
\indent Case D. If $t=\alpha$ or $t=\widehat{b}$, then $f^{*}\in\cal{H}$. For $0<S_{X}(d_{n,n})\leq\ldots\leq S_{X}(d_{n,k})< \alpha < S_{X}(d_{n,k-1}) \leq\ldots\leq S_{X}(d_{n,1})< \widehat{b}$, where $k=2,3,\ldots, n,$
\begin{equation*}
f^{*}(x)=(x-S^{-1}_{X}(\widehat{b}))_{+}\mathbb{I}_{\{m_{2}>0}\}+(x-{\rm VaR}_{\alpha}(X))_{+}\mathbb{I}_{\{m_{2}<0}\}.
\end{equation*}
\indent Case E. If $t=\alpha$ or $t=\widehat{b}$, then $f^{*}\in\cal{H}$. For $\alpha< S_{X}(d_{n,n})\leq\ldots\leq S_{X}(d_{n,l})<\widehat{b}< S_{X}(d_{n,l-1}) \leq\ldots\leq S_{X}(d_{n,1})<1$, where $l=2,3,\ldots, n,$
\begin{equation*}
f^{*}(x)=(x-S^{-1}_{X}(\widehat{b}))_{+}\mathbb{I}_{\{m_{2}>0}\}+ x\mathbb{I}_{\{m_{2}<0}\}.
\end{equation*}
\indent Case F. If $t=\alpha$ or $t=\widehat{b}$, then $f^{*}\in\cal{H}$. For $0< S_{X}(d_{n,n})\leq \ldots \leq S_{X}(d_{n,k})< \alpha < S_{X}(d_{n,k-1}) \leq \ldots \leq S_{X}(d_{n,l})< \widehat{b}< S_{X}(d_{n,l-1})\leq\ldots \leq S_{X}(d_{n,1})<1$, where $l=2,3,\ldots,k-1,$ $k=3,4,\ldots,n,$
\begin{equation*}
f^{*}(x)=(x-S^{-1}_{X}(\widehat{b}))_{+}\mathbb{I}_{\{m_{2}>0}\}+[c_{1}x+c_{2}(x-{\rm VaR}_{\alpha}(X))_{+}]\mathbb{I}_{\{m_{2}<0}\}.
\end{equation*}
where $c_{1}>0,c_{2}>0,$ and $c_{1}x+c_{2}(x-VaR_{\alpha}(X))_{+}\leq x.$
\begin{remark}
\indent Under the $\rm{VaR}$ risk measure, from (4.1) we obtain that $\underline{K}=0,$ $\overline{K}=\frac{1}{\alpha}$ and $K(t)=0$ for $t\in(0,\alpha)$. If $t\in(0,\alpha)$ and $M=0$, then $f^{*}\in\cal{H}$. This means that reinsurance contracts can be any increasing convex function for all $t\in(0,\alpha)$. Moreover, we derive that $S^{-1}_{X}(\widehat{a})={\rm VaR}_{\alpha}(X)$ since $\widehat{a}=\alpha$.
\end{remark}
\begin{remark}
\indent When $\lambda_{1}=\lambda_{2}=\lambda_{3}=0$ and $\beta=1$, our results recover Theorem 3.1 of Cai et al. (2008), Theorem 1 of Cheung (2010) and Theorem 3.1 of Cai and Tan (2010). Moreover, our results are superior to them since optimal reinsurance contracts in our reserch include the combination of quota-share and change-loss and the combination of change-loss and change-loss with different retentions which do not exist in above paper, which means that our finding provides more options for reinsurance strategies.
\end{remark}
\vspace{-1.5em}
\noindent{\bfseries Example 4.1.} Suppose that the ground-up loss $X$ follows an exponential distribution $F_{X}(x)=1-e^{-0.001x}$ for $x\geq 0$, then $E(X)=1000$, ${\rm VaR}_{\alpha}(X)=-1000\ln\alpha$. Let $\alpha=0.05$, then ${\rm VaR}_{\alpha}(X)=2995.73,$ $\frac{1}{\alpha}=20,$ $\underline{K}=0$, $\overline{K}=20.$ Let $\lambda_{1}=0.3,$  $\lambda_{2}=0.4,$ $\lambda_{3}=0.3,$ $\rho=0.2$, then $$M=\frac{(2\beta-0.8)\times 1.2}{2\beta-1.1},$$ if $ m_{2}>0$ and $m_{3}>0$, then $\beta>0.55$; if $m_{2}<0$ and $m_{3}<0$, then $\beta<0.4$. From (4.1) we derive that $\widehat{a}=\alpha=0.05$, so we obtain the following optimal reinsurance.\\
\indent (i) If $m_{2}=0$, then $f^{*}=0$. The case for $m_{2}=0$ is equivalent to $\beta=0.55$, which means it is the optimal option that an insurer undertakes all loss for the case $\beta=0.55$.\\
\indent (ii) When $M\leq0$, if $K(t)=M=0$ for $t\in(0,0.05)$, then $f^{*}(x)\in \cal{H}$, which means that when $\beta=0.4$ and $t\in(0,0.05)$, the optimal reinsurance could be any increasing convex function; for other cases, $f^{*}=0$, for the case of $M<0$, we have $m_{2}<0$ and $m_{3}>0$, which means that when $0.4 < \beta <0.55$, it is the optimal option that an insurer undertakes all loss.\\
\indent (iii) When $M\geq20$, if $K(t^{*})=M=20$ for $t^{*}=0.05$, then $f^{*}(x)\in \cal{H},$ which means that the optimal reinsurance could be any increasing convex function at the point $t=t^{*}$ and $\beta=0.56$; for other cases, $\beta\leq0.56$, moreover, we obtain $f^{*}(x)=0$ for $0.55<\beta<0.56$ and $f^{*}(x)=x$ for $\beta<0.4$.\\
\indent (iv) When $0<M<20$, which is equivalent to $\beta>0.56$, we derive that $f^{*}(x)=0$ for Cases A and C, and $f^{*}(x)=(x-S^{-1}_{X}(\widehat{b}))_{+}$ for Cases B and D-F. Therefore, it is the optimal option that an insurer undertakes all loss for Cases A and C, and the stop-loss reinsurance is optimal for Cases B and D-F. When $t=0.05$ or $t=\widehat{b}$, the optimal reinsurance could be any increasing convex function. $\hfill\square$\\
\vspace{-1.5em}
\subsection{Tail Value at Risk} \setcounter{equation}{1}
\indent Similar to Value at Risk, we consider the case of Tail Value at Risk. Due to the distortion function of TVaR can be described as follows:
\begin{equation*}
g_{\alpha}(t)=\left\{ \begin{aligned}   \overset{} {\frac{t}{\alpha}}, \qquad  t<\alpha, \\ 1 ,  \qquad t\geq \alpha, \end{aligned}   \right.
\end{equation*}
when adopting the expectation premium principle $g_{\gamma}(t)=t$, we have
\begin{equation}
K(t)= {\frac{g_{\alpha}(t)}{g_{\gamma}(t)}}=\left\{ \begin{aligned}   \overset{} {\frac{1}{\alpha}}, \qquad  t<\alpha, \\ \frac{1}{t} ,  \qquad t\geq \alpha. \end{aligned}   \right.
\end{equation}
\vspace{3ex}From Theorem 3.1, we obtain the following proposition.\\
{\bfseries Proposition 4.2.} Assuming the risk is measured by the TVaR risk measure under the expectation premium principle, for $0<\alpha<S_{X}(0)$ and $0<\gamma<S_{X}(0)$, $\beta\in[0,1]$, $\lambda_{i}\geq0$, where $i=1,2,3,$ reinsurance contracts are given as follows:\\
\indent (1) If $m_{2}=0$, then $f^{*}(x)=0$.\\
\indent (2) When $m_{2}\neq0$, considering the following cases.\\
\indent (i) If $M\leq1$, then
\begin{equation*}
f^{*}(x)=x\mathbb{I}_{\{m_{2}>0}\}.
\end{equation*}
\indent (ii) When $M\geq \frac{1}{\alpha}$, if $t\in(0,\alpha]$ and $M=\frac{1}{\alpha}$, then $K(t)=M$, $f^{*}(x)\in \cal{H}.$ For other cases,
\begin{equation*}
f^{*}(x)=x\mathbb{I}_{\{m_{2}<0}\}.
\end{equation*}
\indent (iii) When $1<M<\frac{1}{\alpha}$, we consider the following three cases.\\
\indent Case B. If $t=\widehat{b}$, then $f^{*}\in\cal{H}$. For $0< S_{X}(d_{n,n})\leq S_{X}(d_{n,n-1})\leq \ldots \leq S_{X}(d_{n,1})< \widehat{b}$,
\begin{equation*}
f^{*}(x)=(x-S^{-1}_{X}(\widehat{b}))_{+}\mathbb{I}_{\{m_{2}>0}\}.
\end{equation*}
\indent Case C. If $t=\widehat{b}$, then $f^{*}\in\cal{H}$. For $\widehat{b}< S_{X}(d_{n,n})\leq S_{X}(d_{n,n-1})\leq \ldots \leq S_{X}(d_{n,1})<1$,
\begin{equation*}
f^{*}(x)=x\mathbb{I}_{\{m_{2}<0}\}.
\end{equation*}
\indent Case E. If $t=\widehat{b}$, then $f^{*}\in\cal{H}$. For $0< S_{X}(d_{n,n})\leq\ldots\leq S_{X}(d_{n,l})<\widehat{b}< S_{X}(d_{n,l-1}) \leq\ldots\leq S_{X}(d_{n,1})<1$, where $l=2,3,\ldots, n,$
\begin{equation*}
f^{*}(x)=(x-S^{-1}_{X}(\widehat{b}))_{+}\mathbb{I}_{\{m_{2}>0}\}+ x\mathbb{I}_{\{m_{2}<0}\}.
\end{equation*}
\begin{remark}
\indent Due to $K(t)=\frac{1}{\alpha}$ for $t\in(0,\alpha]$, we derive that $\widehat{a}$ do not exist for the case $\underline{K}<M<\frac{1}{\alpha}$. Therefore, Cases A, D and F do not exist and we only consider above three cases.
\end{remark}
\begin{remark}
\indent When $\lambda_{1}=\lambda_{2}=\lambda_{3}=0$ and $\beta=1$, our results is consistent with Theorem 4.1 of Cai et al. (2008) and Theorem 2 of Cheung (2010).
\end{remark}
\vspace{-1.5em}
\noindent{\bfseries Example 4.2.} Similar to Example 4.1, we consider the case of the TVaR risk measure and obtain the following optimal reinsurance.\\
\indent (i) If $m_{2}=0$, then $f^{*}=0$. The case for $m_{2}=0$ is equivalent to $\beta=0.55$, which means that undertaking all loss for an insurer is the optimal option.\\
\indent (ii) When $M\leq1$, if $m_{2}>0$, then $\beta\leq-0.35$, therefore, there no reinsurance for the case $\beta\in[0,1]$; if $m_{2}<0$ and $m_{3}\leq0$, then $\beta\in[0,0.4],$ $f^{*}=0$; if $m_{2}<0$ and $m_{3}\geq0$, then $\beta\in[0.4,0.55)$, $f^{*}$=0.\\
\indent (iii) When $M\geq20$, if $K(t)=M=20$ for $t\in(0,0.05]$, then $f^{*}(x)\in \cal{H},$ which means that the optimal reinsurance could be any increasing convex function for $t\in(0,0.05]$ and $\beta=0.56$; for other cases, $\beta\leq0.56$, furthermore, we obtain $f^{*}(x)=0$ for $0.55<\beta<0.56$ and $f^{*}(x)=x$ for $\beta<0.4$.\\
\indent (iv) When $1<M<20$, which is equivalent to $\beta>0.56$, we derive that $f^{*}(x)=0$ for Case C, and $f^{*}(x)=(x-S^{-1}_{X}(\widehat{b}))_{+}$ for Cases B and E. Therefore, undertaking all loss for an insurer is the optimal option for Case C, and the stop-loss reinsurance is optimal for Cases B and E. When $t=\widehat{b}$, the optimal reinsurance could be any increasing convex function. $\hfill\square$\\
\vspace{-3em}
\section{Conclusion}
\setcounter{equation}{0}
\indent As we all know reinsurance is an effective risk management tool for the insurer to transfer part of its risk to the reinsurer. However, what we should do is to determine how much risk an insurer should transfer to the reinsurer. This paper discusses two classes of optimal reinsurance models by minimizing their convex combination where the risk is measured by a distortion risk measure and the premium is calculated by a distortion risk premium. We present a unified framework about the unconstrained optimization problem and constrained optimization problems, moreover, not only did we derive the optimal reinsurance strategy but also we derive the minimum of optimization problems by a geometric argument. Under the unified framework, we can derive the solution of the cases from Cases 1-6.

\noindent{\bf\Large Acknowledgements} \ 

\noindent The research was supported by the National Natural Science Foundation of China (No. 11171179, 11571198).


\begin{thebibliography}{10}

\bibitem{A} Arrow, K.J., 1963. Uncertainty and the welfare economics of medical care. The American Economic Review. 53(5), 941-973.

\bibitem{A15} Assa, H., 2015. On optimal reinsurance policy with distortion risk measures and
premiums. Insurance: Mathematics and Economics. 61, 70-75.

\bibitem{B09} Balbás, A., Balbás, B., Balbás, R., Heras, A., 2009. Optimal reinsurance with general
risk measures. Insurance: Mathematics and Economics. 44, 374-384.

\bibitem{B} Borch, K., 1960. An attempt to determine the optimum amount of stop loss
reinsurance. In: Transactions of the 16th International Congress of Actuaries.

\bibitem{CT07} Cai, J., Tan, K.S., 2007. Optimal retention for a stop-loss reinsurance under the VaR
and CTE risk measures. Astin Bulletin. 37(1), 93-112.

\bibitem{CTWZ08} Cai, J., Tan, K.S., Weng, C., Zhang, Y., 2008. Optimal reinsurance under VaR and CTE
risk measures. Insurance: Mathematics and Economics. 43, 185-196.

\bibitem{C10} Cheung, K.C., 2010. Optimal reinsurance revisited: a geometric approach. Astin
Bulletin. 40, 221-239.

\bibitem{CL15} Cheung, K.C., Lo, A., 2017. Characterizations of optimal reinsurance treaties: a cost benefit approach. Scandinavian Actuarial Journal. 1, 1-28.

\bibitem{CSYY14} Cheung, K.C., Sung, K., Yam, S., Yung, S., 2014. Optimal reinsurance under general
law-invariant risk measures. Scandinavian Actuarial Journal. 72-91.

\bibitem{CT11} Chi, Y., Tan, K.S., 2011. Optimal reinsurance under VaR and CVaR risk measures: a
simplified approach. Astin Bulletin 41(2), 487-509.

\bibitem{CT13} Chi, Y., Tan, K.S., 2013. Optimal reinsurance with general premium principles.
Insurance: Mathematics and Economics. 52, 180-189.

\bibitem{CW13} Chi, Y., Weng, C., 2013. Optimal reinsurance subject to Vajda condition. Insurance:
Mathematics and Economics. 53, 170-189.

\bibitem{CYW13} Cui, W., Yang, J., Wu, L., 2013. Optimal reinsurance minimizing the distortion
risk measure under general reinsurance premium principles. Insurance: Mathematics and Economics. 53, 74-85.

\bibitem{DKLT12} Dhaene, J., Kukush, A., Linders, D., Tang, Q., 2012. Remarks on quantiles and
distortion risk measures. European Actuarial Journal. 2, 319-328.

\bibitem{JHR} Jiang, W., Hong, H., Ren, J., 2017. On Pareto-optimal reinsurance with constraints under distortion risk measures. European Actuarial Journal. 8(1), 215-243.

\bibitem{K01} Kaluszka, M., 2001. Optimal reinsurance under mean-variance premium principles.
Insurance: Mathematics and Economics. 28, 61-67.

\bibitem{KO08} Kaluszka, M., Okolewski, A., 2008. An extension of Arrow’s result on optimal reinsurance
contract. The Journal of Risk and Insurance. 75, 275-288.

\bibitem{LZY2015} Li, P., Zhou, M., Yin, C.C., 2015. Optimal reinsurance with both proportional and fixed costs. Statistics and Probability Letters. 106, 134-141.

\bibitem{TWZ11} Tan, K.S., Weng, C., Zhang, Y., 2011. Optimality of general reinsurance contracts
under CTE risk measure. Insurance: Mathematics and Economics. 49, 175-187.

\bibitem{Y} Young, V.R., 1999. Optimal insurance under Wang’s premium principle. Insurance: Mathematics and Economics. 25, 109-122.

\bibitem{Y2018} Yin, C.C., 2018. Remarks on Equality of two distributions under some partial orders. Acta Mathematicae Applicatae Sinica, English Series. 34(2), 274-280.

\bibitem{YZ2016} Yin, C.C., Zhu, D., 2016. New class of distortion risk measure and their tail asymptotics with emphasis on VaR. Journal of Financial Risk Management. 7, 12-23.

\bibitem{ZC2014} Zheng, Y.T., Cui, W., 2014. Optimal reinsurance with premium constraint under
distortion risk measures. Insurance: Mathematics and Economics. 59, 109-120.

\bibitem{ZWTA2016} Zhuang, S.C., Weng, C., Tan, K.S., Assa, H., 2016. Marginal indemnification function
formulation for optimal reinsurance. Insurance: Mathematics and Economics. 67, 65-76.

\end{thebibliography}
\end{document}